# An ideal independent source as an equivalent 1-port

*Emanuel Gluskin* [1]∗  and  *Anatoly Patlakh* [2]

1. Electrical Engineering Departments of the ORT Braude College, the Kinneret Academic College on the Sea of Galilee, and Ben-Gurion University of the Negev, Beer-Sheva 84105.

2. Electrical Engineering Department., Afeka Academic College,
218 Bni Efraim St., Tel Aviv 69107.

∗ *Communicating author*:  gluskin@ee.bgu.ac.il , http://www.ee.bgu.ac.il/~gluskin/

**Abstract**:  We consider a 1-port, not necessarily linear, with a dependent source, appearing at the port.  The control of the source is entirely internal for the 1-port.  If this source is a parallel voltage source, then the equivalent circuit is an ideal independent voltage source, and if it is a series current source, then the equivalent circuit is an ideal independent current source. (As usual, "ideal" source is defined as a source whose proposed function is independent of the load.)  In the simple LTI case, these results can be obtained, respectively, by either taking $R_{Th}$ zero in the Thevenin equivalent, or taking $R_N$ infinite in the Norton equivalent; however the very fact that the final circuits do not include any linear elements indicates the possibility of generalization to nonlinear 1-ports.  Some limitations on the circuit's structure (functional dependencies in it) are required, and the clearness of these limitations, i.e. clearness of the conditions for the 1-port to be an ideal source for any load, is the aesthetical point.

## 1.  Introduction

Simplifications of linear circuits obtained by using Thevenin's theorem (the *series $v_{Th}$-$R_{Th}$* circuit), or Norton's equivalent (the *parallel $i_N$||$R_N$* circuit) are very useful in many problems [1-3].  As far as we speak about linear circuits, the two theorems proved below can be seen as closely associated with Thevenin's and Norton's theorems.  However, while for the actual circuits that are relevant to the known theorems, the particular case when $v_{Th} = 0$ (or $i_N = 0$) is often obtained [2], in our situation we obtain a very rare "opposite" case of $R_{Th} = 0$ (or $R_N = \infty$).  The usual passing, in the linear case, from the Thevenin circuit to the Norton circuit and back can *not* be done here on the basis of the equivalent generator theorem in which the equality $i_N = v_{Th}/R_{Th}$ is used, and each of the circuits arising here has only one equivalent, either of Thevenin, or of Norton, i.e. the sources shown in Figs. 2 and 6 have to be derived independently, or using the duality argument.

As far as only linear circuits are concerned, the results are also relevant for "impedance" circuits formulated in the $\omega$- and $s$- (Laplace variable) domains.

The situation as regards the separation of the two schemes discussed, one with a current and the other with a voltage source, continues in the much more interesting nonlinear case, which is the general one in Section 2.  (The linear case is included as a particular case.)  No reason is seen to exclude chaotic circuits, even though no reasons for chaos can be analyzed here.

## 2.  The circuit with an "output" dependent voltage source as an independent ideal voltage source

Consider the schematic 1-port in which the specified control of the output dependent voltage source is internal for this 1-port, not including any externally generated voltage or current.

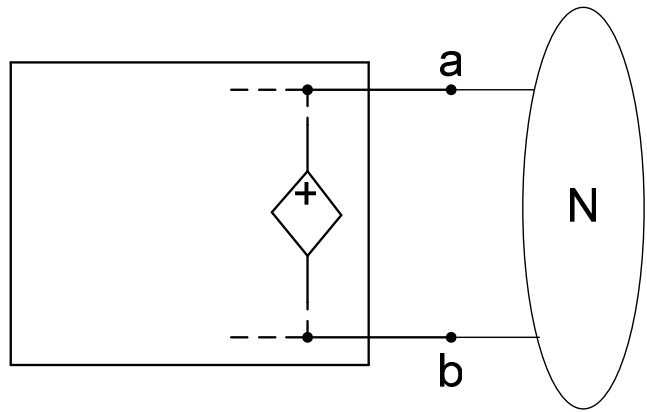

Fig. 1: The general circuit for Theorem 1. One can require the 1-port (generally including some independent sources) to the left from **a-b** to have, when detached, a unique solution, which is relevant to engineering applications.  However, the main point is only that the control of the dependent source is *internal*, belonging to the 1-port. Node **b** is *grounded*, and the potential $v_a$ of node **a** (i.e. $v_{ab}$) is of especial interest.  That this potential is functionally given via the characteristic of the source is the point of the proof of Theorem 1. Because of the output voltage source, one can not connect an arbitrary voltage source ("$v_{test}$") to the port, or short-circuit it, but these limitations are the only ones on external circuit N.

**Theorem 1**: *With the limitation on the circuit structure noted in Section 4, the circuit of Fig. 1, as seen from the port, is equivalent to the circuit shown in Fig. 2, i.e. to an independent voltage source.*

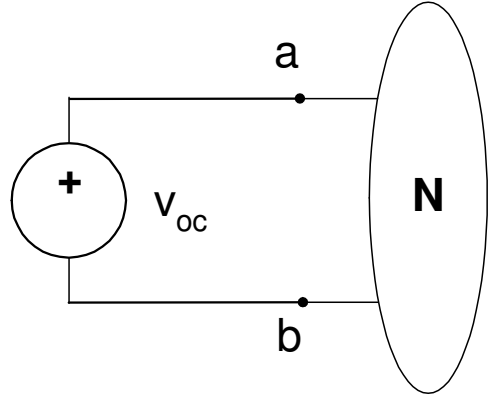

Fig. 2: The equivalent to the circuit in Fig. 1. *In the linear case*, it can be seen as "just" a case of Thevenin's theorem for $R_{Th} = 0$, but a point is that this equivalent can be relevant also to a nonlinear 1-port.

The physical sense of the fact that $i_{test}$ (or the entering current) does not influence $v_{ab}$ is that the ideal voltage source can absorb an unlimited current. The latter is not associated with the possibility of linear control of the source.

The main point is to see whether or not the input current of the 1-port, caused by the external connection, can influence the variables controlling the dependent source, which all belong to the 1–port. We argue that *the output dependent source defends all of its controls, and thus also itself*, *from any external influence*.

Proof of Theorem 1:

Let us use nodal voltages analysis, -- a general circuit method [2] which gives a complete circuit description. Using this method, we write KCL equations for branch currents expressed via nodal voltages. However if a voltage source (even a dependent one) is directly connected between two adjacent nodes, then no KCL equation is needed for one of these nodes, because its voltage is already equationally expressed via the other nodal voltage, according to the specified function of the source.

We formalize this by writing for node **a**

$$v_a = f(x_1, x_2, ..., x_p) \qquad (1)$$

where $x_1, x_2, ..., x_p$ are the *p* variables that control the output source (below "the controls"). It can be that among these controls there are some of the nodal voltages, our main unknowns.

It is realistic to require that for all *p* arguments zero,

$$f(0, 0, ..., 0) = 0 . \qquad (2)$$

We assume, furthermore, that it is possible to present $x_1, x_2, ..., x_p$ via some known functions $\xi_1, \xi_2, ..., \xi_p$

$$x_1 = \xi_1(v_1, v_2, ..., v_n)$$
$$x_2 = \xi_2(v_1, v_2, ..., v_n)$$
$$..............................$$
$$x_p = \xi_p(v_1, v_2, ..., v_n)$$

(3)

where the arguments are all of the nodal voltages, while $v_n = v_a$. Substituting (3) into (1), we obtain

$$v_a = F(v_1, v_2, ..., v_{n-1}, v_a), \quad (4)$$

and after resolving for $v_a$,

$$v_a = \psi(v_1, v_2, ..., v_{n-1}) \quad (4a)$$

with a known function $\psi$.

It is important for the point that no other equation for node **a** is needed, because we have, in total, enough equations, and each additional one cannot be independent. Indeed, using the incident matrix of the circuit [1], we express, furthermore, the branch currents involved in the *n-1* KCL equations written for the *n-1* internal nodes, via all the nodal voltages (including $v_a$), having, together with (4), *n* equations for the *n* voltages.

Consider two possible cases of the circuit topology. The first case is when there are no nodes in the 1-port directly connected to node **a**, i.e. apart of its high-impedance control inputs, the output dependent source just has a common ground with the rest of the circuit. Then, we have an independent nodal voltages problem for all the other nodes, and, using the KCL equations, find $v_1, v_2, ..., v_{n-1}$ independently from the output source. Then, (4) gives $v_a$.

The other possibility is that there are some nodes directly (via simple branches) connected to **a**. Then voltage $v_n = v_a$ is included in the currents (and only in these currents) of the branches entering **a**, and appear in the KCL equations for the nodes connected to **a**.

In this case, we first use (4) in order to eliminate $v_a$ from the latter equations by substituting into them $v_a = \psi(v_1, v_2, ..., v_{n-1})$, thus obtaining *n*-1 equations including only the unknowns $v_1, v_2, ..., v_{n-1}$. Thus $v_1, v_2, ..., v_{n-1}$ can be found, and then $v_a = \psi(v_1, v_2, ..., v_{n-1})$ becomes known too.

We thus have a complete circuit solution, and there is no place in this procedure for the possible external current to influence anything in the 1-port, including $v_a$, i.e. the 1-port is as a whole is an ideal voltage source independent of the outer conditions.

Thus, for any solvable circuit that can be solved in such a way, Theorem 1 is proved.

Comment 1:   Since our 1-port can not be influenced from outside, if we "switch of" the internal independent sources, we must obtain (in agreement with (2)) the output voltage zero, i.e. the equivalent source zeroed.

Comment 2:   If we deal with a chaotic circuit, then the obtained independent source can generate a chaotic function (process) independent of the external conditions.

## 3. Examples

Two examples, one for a linear 1-port, and one for a nonlinear 1-port, follow. In both cases the point is that despite the external connection, the output voltage is defined only by the internal elements of the 1-port.

### 3.1. *Linear circuit*

Consider the circuit in Fig. 3

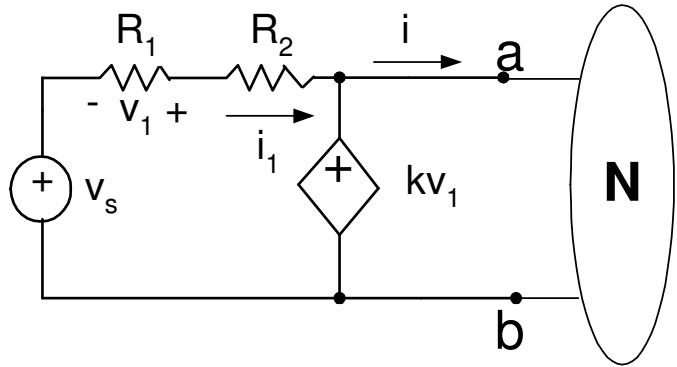

Fig. 3:   An example for Theorem 1. Voltage $v_{ab}$ is independent of $i$ and the circuit N. The whole circuit to the left from **a-b** appears to be an ideal independent source.

KVL over the left mesh gives:

$$v_{ab} = kv_1 = -kR_1 i_1 = -kR_1 \frac{v_s + (1/k)v_{ab} - v_{ab}}{R_2} \qquad (5)$$

(we used that $v_1 = (1/k)v_{ab}$), from which

$$v_{ab} = \frac{kR_1}{(k-1)R_1 - R_2} v_s$$

which is independent of $i$ and the external circuit. This is, of course, also the open-circuit voltage $v_{oc}$ (i.e. the special case of $i = 0$) of the 1-port to the left of **a-b**.

### 3.2. *Nonlinear circuit*

To create a nonlinear circuit, let us replace, in the previous circuit (Fig.3), the linear resistor $R_2$ by the nonlinear resistor given by the relation

$$v_{R_2} = \alpha i^2, \quad \alpha > 0, \quad (i = i_1) \qquad (6)$$

(now "$R_2$" is just the specification *name* of this element), i.e., using the arithmetic (positive) value of the square root,

$$i_1 = \pm v_{R_2} \sqrt{\frac{v_{R_2}}{\alpha}} .$$

Then, (5) becomes

$$v_{ab} = kv_1 = -kR_1 i_1 = \mp kR_1 \sqrt{\frac{v_s + (1/k)v_{ab} - v_{ab}}{\alpha}} . \qquad (7)$$

After solving a second-order algebraic equation for $v_{ab}$, obtained from (7), $v_{ab}$ is finally expressed via $v_s$ as

$$v_{ab} = \frac{kR_1^2}{2\alpha}\left(1 - k \pm \sqrt{(1-k)^2 + 4\frac{\alpha}{R_1^2}v_s}\right) .$$

We see that here too the external (load's) current is not influencing.
A somewhat simpler, strongly nonlinear example is obtained using the model

$$v_{R_2} = R|i|, \quad R > 0, \quad (i = i_1) \qquad (6a)$$

instead of (6), in the same circuit. Then

$$i = \pm \frac{|v_{R_2}|}{R} ,$$

and (5) becomes

$$v_{ab} = kv_1 = -kR_1 i_1 = \mp \frac{kR_1}{R}|v_s + (1/k)v_{ab} - v_{ab}| .$$

Requiring intersection of the function of the type $y = |x-a|$ with the function of the type $y = x$, one easily graphically sees that for a solution $v_{ab}$ to exist we have to take before the right-hand side of the latter equation '+' for $k > 1$ (then $v_{ab} > 0$), and '-' for $k < 1$ (then $v_{ab} < 0$).  In any case, $v_{ab}$ depends only on the internal circuit parameters.

## 4. A limitation on the circuit's internal control dependencies

For the circuits relevant to Theorem 1, it is necessary to require that the circuit *does not include any internal dependent source for which* $v_a$, *or the current flowing through the output dependent voltage source*, *is one of its controls*. The following example which employs such current control, contradicts Theorem 1. The obvious reason for the problem is that the current of the voltage source , $i_1$, is dependent on the load's current, and the load is allowed to influence $v_{ab}$ via the current feedback.

Simple calculations show that for this circuit

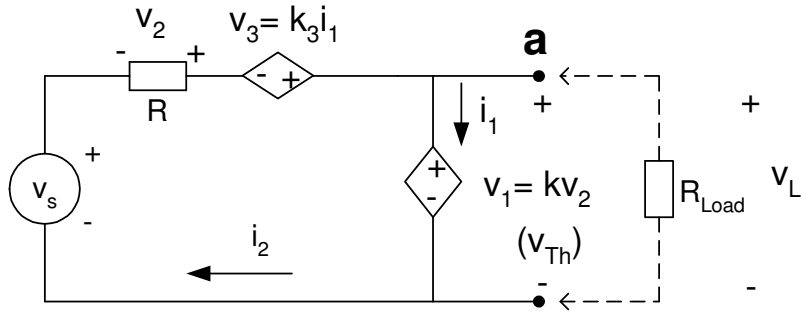

Fig. 4: The circuit with an internal dependent source, influenced (controlled) by the current of the output dependent voltage source. In this case, $v_{Th} \neq v_L$, which states a limitation on the conditions of applicability of Theorem 1. Observe for (8) and (9) that the physical dimension $[k_3] = \Omega$.

$$v_{Th} = \frac{v_s}{\frac{k_3}{kR} - \frac{1}{k} + 1}, \tag{8}$$

but

$$v_{Load} = \frac{v_s}{k_3(\frac{1}{kR} + \frac{1}{R_{Load}}) - \frac{1}{k} + 1}. \tag{9}$$

Thus here $v_L$ depends on $R_{Load}$ and $v_{Th} \neq v_L$, contradicting Theorem 1.

The cases of such "back-influence" of the current of the output dependent voltage source are thus irrelevant to Theorem 1.

## 5. The case of output current source

Consider now the circuit (that also need not be linear) shown in Fig. 5, with a dependent output current source.

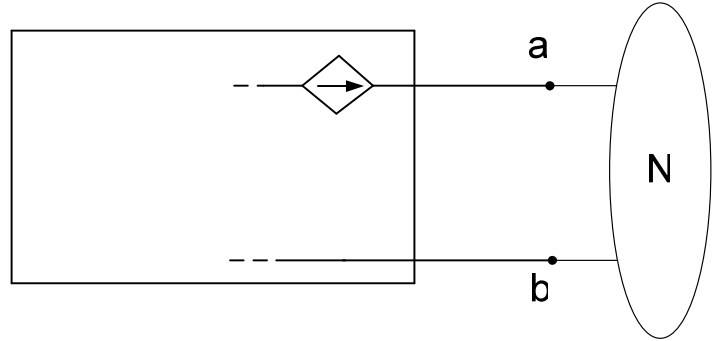

Fig. 5: The circuit to be simplified to an ideal current source. The output current of the 1-port is of especial interest. We can connect a voltage-test source, $v_{test}$, thus defining $v_{ab}$, but can neither leave the circuit open, nor connect any series source of $i_{test}$. These are the only limitations to the external circuit N. The point is to show that $v_{ab}$ does not influence the output current of the 1-port.

Of course, such a circuit cannot be detached from the external circuit, but we can replace the external circuit by some $v_{test}$, in particular, short-circuit it.

**Theorem 2**: *With the limitations on the circuit structure, or its internal functional connections, dual to those of Theorem 1 (i.e. $v_{ab}$ may not influence any internal dependent source, compare to Section 4), the equivalent of the circuit of Fig. 5 is the circuit of Fig. 6, i.e. the 1-port is an ideal independent current source.*

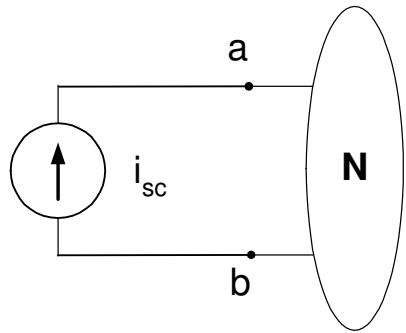

Fig. 6: The equivalent of the circuit in Fig.5.

Since the equivalent-generator transformation is not possible in the relevant circuits, we cannot obtain this result automatically from Theorem 1, but can use concepts that are *dual* with respect to those of Theorem 1, i.e. the mesh currents instead of nodal voltages, and the associated KVL equations instead of the KCL equations. Then, the steps of the proofs of Theorem 1 are essentially repeated. In particular, we use now *mesh-current* analysis, expressing the nodal voltages via the mesh currents as the unknowns. Then, at the output loop, we have that the mesh current, -- *which is the current of the dependent source*, -- is already expressed via the internal state-variables, so that no KVL equation is needed for this mesh and thus $v_{test}$ can *not* be equationally included. We finally conclude that the output current is only due to the internal independent sources, i.e. $i_{out} = i_{sc}$ (short-circuit current) and the circuit of Fig. 6 indeed is equivalent to the circuit of Fig. 5.

Of course, nonlinear circuit versions can be relevant here too.

The physical reason for the result proved is the ability of an ideal current source withstand any voltage drop, and it is just prohibited to use this voltage drop as a control for any internal dependent source.

## Conclusions

1. Apart from the often obtained case when Thevenin (or Norton) equivalent circuit is just a passive resistor, the opposite case when $v_{Th}$, or $i_N$ is nonzero, but the resistor is of zero (or infinite) value, can be observed, and is worth considering. This case should be useful in creating of the "ideal" controllable voltage or current sources.

2. The equivalent-generator theorem is irrelevant, i.e. the Thevenin and Norton cases are "split" here, but the duality argument can be applied.

3. The absence of $R_{Th}$ or $R_N$ in such an equivalent circuit makes the proved theorem(s) relevant to nonlinear internal structure of the source, and chaotic circuits that would compose "ideal" generators of chaotic processes should not be excluded. The possible relevance to nonlinear circuit design might be an important completion to the classical Thevenin-Norton applicational scheme.

4. Construction of an ideal, itself not chaotic source feeding an *external* chaotic circuit, can also be of an interest here, because of the reasons given in [4] and references mentioned there.

5. For a linear circuit, the "algebraic" proofs of the given theorems can be relevant also to the "impedance"-type circuits in the *s*- and $\omega$- domains.

6. For Theorem 1, current feedback from the terminating source is prohibited, and for Theorem 2, voltage feedback is prohibited. These simple and clearly justified (almost directly following from the very definition of the independent source) limitations are the only limitations on application of the proved theorems. The clarity of these limitations should make the theorems simple in use, and we find this clarity to be the aesthetic point of the circuit situation.

7. Generally, the functional *control* dependences in the source's circuit cannot be converted, which means that if an *internal* dependent source is controlled by the output (the terminating) dependent source, Theorem 1 or 2 can fail.

8. The classical Thevenin/Norton theorem still includes interesting research aspects to which the attention of the electronics and circuits specialists should be drawn. The whole research scheme obviously includes many possibilities for computer (and PSpice, etc.) simulations, and a teacher can find here a good material for students' projects. Especially interesting would be cascade connection of such 1-ports (e.g., the source $v_s$ in Fig. 3 is the previous 1-port), and also creation of such chaotic circuits. The proved theorems should be included in the text-books on basic circuit theory.

9. The circuits with *terminating* dependent sources are often met in the theory of amplifiers. We think, however, that such examples should be treated also at the much earlier educational stage when the basic Thevenin theorem is introduced, and the expected circuitry is much more general.

**Acknowledgement**:  We are grateful to Ralf Lehnert for some helpful comments.